\documentclass[
 reprint,twocolumn,
 amsmath,amssymb,
 aps,
]{revtex4}

\usepackage{epsfig,bm,epsf,graphics}
\usepackage{graphicx}
\usepackage{dcolumn}
\usepackage{bm}

\begin{document}
\preprint{APS/123-QED}

\title{Helimagnetic thin films: surface reconstruction, surface spin-waves and magnetization}

\author{Sahbi El Hog\footnote{sahbi.el-hog@u-cergy.fr} and H. T. Diep\footnote{diep@u-cergy.fr}}
 \affiliation{%
Laboratoire de Physique Th\'eorique et Mod\'elisation,
Universit\'e de Cergy-Pontoise, CNRS, UMR 8089\\
2, Avenue Adolphe Chauvin, 95302 Cergy-Pontoise Cedex, France.\\
 }%





\begin{abstract}

Quantum properties of a helimagnetic thin film of simple cubic lattice with
Heisenberg spin model are studied using the Green's function method.
We find that the spin configuration across the film
is strongly non uniform.
Using the exactly determined spin configuration we calculate  the
spin-wave spectrum and the layer magnetizations as  functions of temperature $T$.
We show the existence of surface-localized modes which strongly affect the surface magnetization.
We also show that quantum fluctuations cause interesting spin contractions at $T=0$ and give rise to a cross-over between layer magnetizations at low $T$.
\begin{description}
\item[PACS numbers: 75.25.-j ; 75.30.Ds ; 75.70.-i ]
\end{description}
\end{abstract}

\pacs{Valid PACS appear here}
\maketitle


\section{Introduction}
Recently, there has been a growing interest in magnetic properties of helimagnets, due to possible applications in spin transport properties using materials at nanoscale such as thin films and multilayers \cite{Fert2013}.  In particular, intensive researches have been carried out to understand the role of skyrmions \cite{Lin,Bogdanov,Rossler,Muhlbauer,Yu1,Yu2,Seki,Adams}.
There is in addition a large number of experiments which has recently been performed on thin films of helical magnets \cite{Heurich,Wessely,Jonietz}.

The simplest model of
the helimagnetic ordering
is due to a competition between nearest-neighbor (NN)  and next-nearest-neighbor (NNN) interactions, as discovered by Yoshimori \cite{Yoshimori} and Villain \cite{Villain59}: a spin in a chain turns an angle $\theta$ with respect to its previous neighbor.  There are many families of helimagnets due to various kinds of interaction among them one can mention non collinear magnetic structures due to Dzyaloshinskii-Moriya interactions or to geometry frustration \cite{Bak,Plumer,Maleyev}.
Low-temperature properties in helimagnets such as spin-waves \cite{Harada,Rastelli,Diep89,Quartu1998} and
heat capacity \cite{Stishov} have been extensively investigated. In spite of their long history, the nature of the phase transition in non collinear magnets such as stacked
triangular XY and Heisenberg antiferromagnets has been elucidated only recently \cite{Diep89b,Ngo08,Ngo09}. For reviews, the reader is referred to Ref. \onlinecite{DiepFSS}.

In this paper, we study a quantum Heisenberg helimagnetic thin film with the simple cubic (sc) lattice. The case of the body-centered cubic (bcc) lattice has been recently studied \cite{Diep2015}.
Surface effects in thin films have
been intensively studied during the last three decades \cite{Heinrich,Zangwill}. However, due to complicated surface spin configurations, surface effects in helimagnets have
only been recently studied: surface spin structures \cite{Mello2003}, Monte Carlo (MC) simulations \cite{Cinti2008} and a few experiments \cite{Karhu2011,Karhu2012}. Helical magnets present potential applications in spintronics with predictions of spin-dependent electron transport in these magnetic materials \cite{Heurich,Wessely,Jonietz}.  This motivates the present work.

We shall use the Green's function (GF) method which has been initiated
by Diep-The-Hung {\it et al.} for collinear
surface spin configurations \cite{Diep1979}.  For non collinear magnets, the GF method has also been developed for bulk
helimagnets \cite{Quartu1998} and for frustrated films \cite{NgoSurface,NgoSurface2}.  In helimagnets, the angles between neighboring spins become strongly non uniform as seen below, making calculations harder. This explains the small number of microscopic calculations so far for helimagnetic films.

The paper is organized as follows. In section II, the model is presented and classical ground state (GS) of the helimagnetic film is determined. We summarize there the principal steps used in the general GF method for non-uniform spin configurations. The GF results are shown in section III where the spin-wave spectrum, the zero-point spin contraction and the layer magnetizations are shown. Concluding remarks are given in section IV.

\section{Model, classical ground state and quantum formulation}\label{GSSC}

We consider a thin film of sc lattice of $N_z$ layers, with two symmetrical surfaces perpendicular to
the $c$-axis, for simplicity.  The exchange Hamiltonian is given by
\begin{equation}
\mathcal H_e=-\sum_{\left<i,j\right>}J_{i,j}\mathbf S_i\cdot\mathbf
S_j  \label{eqn:hamil1}
\end{equation}
where $J_{i,j}$ is the interaction between two quantum Heisenberg spins $\mathbf S_i$ and $\mathbf S_j$ occupying the lattice sites $i$ and $j$.

\subsection{Surface spin reconstruction}
To generate a bulk helimagnetic structure, the simplest way is to take a ferromagnetic interaction between NNs  $J_1$ ($>0$),
and an antiferromagnetic interaction between NNNs  $J_2<0$.  If $|J_2|$ is smaller than a critical value $|J_2^c|$,
the classical GS spin configuration is ferromagnetic \cite{Harada,Rastelli,Diep89}.
Let us consider the case of a
helimagnetic structure only in the $c$-direction perpendicular to the film surface. In such a case, we assume a non-zero $J_2$ only on the $c$-axis.
This assumption simplifies formulas but does not change the physics of the problem since including the uniform helical angles in two other
directions parallel to the surface will not introduce additional surface effects.  The bulk quantum helimagnets have been studied by the Green function method \cite{Quartu1998}.

 For the present model, the helical structure in the bulk is planar: spins are parallel in planes perpendicular to the $c$-axis and the angle between two NNs in
the adjacent planes is a constant and is given by $\cos \alpha=-\frac{J_1}{4J_2}$ for a sc lattice. The helical structure exists therefore if $|J_2|> 0.25 J_1$,
namely $|J_2^c|$(bulk)$=0.25 J_1$. To calculate the classical GS surface spin configuration, we write down the expression of the energy of spins along the $c$-axis, starting from the surface:
\begin{eqnarray}
E&=& -J_1 \cos (\theta_1-\theta_2)-J_1 [\cos (\theta_2-\theta_1)\nonumber\\
&&+ \cos (\theta_2-\theta_3)]+...\nonumber\\
&&-J_2 \cos (\theta_1-\theta_3)-J_2 \cos (\theta_2-\theta_4)\nonumber\\
&&-J_2[\cos (\theta_3-\theta_1)+ \cos (\theta_3-\theta_5)]+...\label{EC}
\end{eqnarray}
where  $\theta_i$ denotes the angle of a spin in the $i$-th layer
made with the Cartesian $x$ axis of the layer. The interaction energy between two NN spins in the two adjacent layers $i$ and $j$
depends only on the difference $\alpha_{i}\equiv \theta_i-\theta_{i+ 1}$. The GS configuration corresponds to the minimum of $E$.
We have to solve by iteration the set of equations:
\begin{equation}
\frac{\partial E}{\partial \alpha_i}=0, \ \ \ \mbox{for}\ \ i=1,N_z-1
\end{equation}

\begin{figure}[ht!]
\centering
\includegraphics[width=7cm,angle=0]{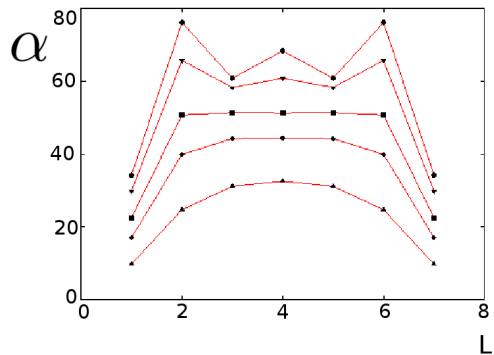}
\caption{(Color online) Angles $\alpha_1 ...... \alpha_7$ in degree across the film for $J_2$=-0.6, -0.5, -0.4, -0.35, -0.3 (from top) with $N_Z=8$.\label{angle}}
\end{figure}

The result is shown in Fig. \ref{angle} for $N_z=8$. Some remarks are in order: i) the result is obtained by iteration with errors less than $10^{-4}$ degrees,  ii) strong angle variations  are observed near the surface with oscillation for strong $|J_2|$, iii) the angles at the film center are close to the bulk value $\alpha$, meaning that the surface reconstruction affects just a few atomic layers (this is more clearly seen for thicker films not shown here). This bulk helical stability has been experimentally observed in holmium films \cite{Leiner}.
An alternative method giving the same result is the numerical steepest descent method which is described in details in Ref. \onlinecite{NgoSurface}.


\subsection{Analytical formulation}

\begin{figure}[htb]
\centering
\includegraphics[width=7cm]{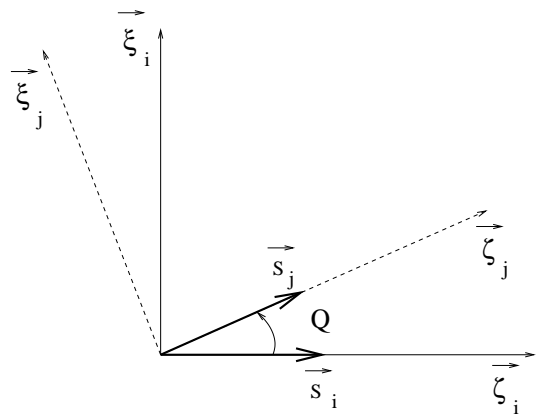}  
\caption{ Local coordinates in a $xy$-plane perpendicular to the $c$-axis. $Q$ denotes $\theta_{j}-\theta_i$.}\label{local}
\end{figure}

To calculate physical quantities at finite temperatures, we shall use the GF method. To that end, we use the local spin coordinates defined as follows \cite{Quartu1998,Diep2015}:  the quantization axis of spin $\vec S_i$ is on
its $\zeta_i$ axis which lies in the plane, the $\eta_i$ axis of $\vec S_i$ is along the $c$-axis, and the $\xi_i$
axis forms with $\eta_i$ and $\zeta_i$ axes a direct trihedron (see Fig. \ref{local}).

Expressing the Hamiltonian in the local coordinates, we obtain
\begin{eqnarray}
\mathcal H_e &=& - \sum_{<i,j>}
J_{i,j}\Bigg\{\frac{1}{4}\left(\cos\theta_{ij} -1\right)
\left(S^+_iS^+_j +S^-_iS^-_j\right)\nonumber\\
&+& \frac{1}{4}\left(\cos\theta_{ij} +1\right) \left(S^+_iS^-_j
+S^-_iS^+_j\right)\nonumber\\
&+&\frac{1}{2}\sin\theta_{ij}\left(S^+_i +S^-_i\right)S^z_j
-\frac{1}{2}\sin\theta_{ij}S^z_i\left(S^+_j
+S^-_j\right)\nonumber\\
&+&\cos\theta_{ij}S^z_iS^z_j\Bigg\}
\label{eq:HGH2}
\end{eqnarray}
Now, according to the theorem of Mermin and Wagner \cite{Mermin}
continuous isotropic spin models such as XY and Heisenberg spins
do not have long-range ordering at finite temperatures in two dimensions. Since our films have small thickness, it is useful to add an anisotropic interaction to stabilize the long-range ordering at finite temperatures.
Let us use the following in-plane anisotropy between $\mathbf S_i$
and $\mathbf S_j$:
\begin{equation}
\mathcal H_a= -\sum_{<i,j>} I_{i,j}S^z_iS^z_j\cos\theta_{ij}
\end{equation}
where $I_{i,j}(>0)$ is supposed to be positive, small compared to $J_1$, and limited to NNs.
The full Hamiltonian is thus
$\mathcal H=\mathcal H_e+\mathcal H_a$. The GS in the presence of $I_{i,j}(>0)$ can be determined in the same manner. Hereafter we take $I_{i,j}=I_1$ for any NN pair, except otherwise stated. It is only very slightly modified with the order of one or two degrees when $I_1\simeq 0.1J_1$. The small anisotropy  does not therefore alter the main features shown in Fig. \ref{angle}.


The general method has been recently described in details in Ref. \onlinecite{Diep2015}. To save space, let us just briefly recall here the principal steps of calculation and give the results for the sc helimagnetic film only where they should be.
We define the following two double-time Green's functions in the real space:
\begin{eqnarray}
G_{i,j}(t,t')&=&<<S_i^+(t);S_{j}^-(t')>>\nonumber\\
&=&-i\theta (t-t')
<\left[S_i^+(t),S_{j}^-(t')\right]> \label{green59a}\\
F_{i,j}(t,t')&=&<<S_i^-(t);S_{j}^-(t')>>\nonumber\\
&=&-i\theta (t-t')
<\left[S_i^-(t),S_{j}^-(t')\right]>\label{green60}
\end{eqnarray}
We need these two functions because the equation of motion of the first function generates functions of
the second type, and vice-versa. Writing the equations of motion of these functions
and using the Tyablikov decoupling scheme to reduce the
higher-order functions, for example $<<S_{i'}^zS_i^+(t);S_{j}^-(t')>>\simeq <S_{i'}^z><<S_i^+(t);S_{j}^-(t')>>$ etc., we
obtain the general equations for non collinear magnets \cite{Diep2015}.
We next introduce the following in-plane Fourier
transforms:

\begin{eqnarray}
G_{i, j}\left( t, t'\right) &=& \frac {1}{\Delta}\int\int_{BZ} d\mathbf
k_{xy}\frac{1}{2\pi}\int^{+\infty}_{-\infty}d\omega e^{-i\omega
\left(t-t'\right)}\nonumber\\
&&\hspace{0.7cm}\times g_{n_i,n_j}\left(\omega , \mathbf k_{xy}\right)
e^{i\mathbf k_{xy}\cdot \left(\mathbf R_i-\mathbf
R_j\right)},\label{eq:HGFourG}\\
F_{k, j}\left( t, t'\right) &=& \frac {1}{\Delta}\int\int_{BZ} d\mathbf
k_{xy}\frac{1}{2\pi}\int^{+\infty}_{-\infty}d\omega e^{-i\omega
\left(t-t'\right)}\nonumber\\
&&\hspace{0.7cm}\times f_{n_k,n_j}\left(\omega , \mathbf k_{xy}\right)
e^{i\mathbf k_{xy}\cdot \left(\mathbf R_k-\mathbf
R_j\right)},\label{eq:HGFourF}
\end{eqnarray}
where $\omega$ is the spin-wave frequency, $\mathbf k_{xy}$
denotes the wave-vector parallel to $xy$ planes and $\mathbf R_i$ is
the position of the spin at the site $i$. $n_i$, $n_j$ and $n_k$ are
respectively the $z$-component indices of the layers where the sites $\mathbf R_i$,  $\mathbf R_j$ and $\mathbf R_k$
belong to. The integral over $\mathbf k_{xy}$ is performed in the
first Brillouin zone ($BZ$) whose surface is $\Delta$ in the $xy$
reciprocal plane.  For convenience, we denote $n_i=1$ for all sites on the surface layer,
$n_i=2$ for all sites of the second layer and so on.

We finally obtain the following matrix equation
\begin{equation}
\mathbf M \left( \omega \right) \mathbf h = \mathbf u,
\label{eq:HGMatrix}
\end{equation}
where $\mathbf M\left(\omega\right)$ is a square matrix of dimension
$\left(2N_z \times 2N_z\right)$, $\mathbf h$ and $\mathbf u$ are
the column matrices which are defined as follows
\begin{equation}
\mathbf h = \left(%
\begin{array}{c}
  g_{1,n'} \\
  f_{1,n'} \\
  \vdots \\
  g_{n,n'} \\
  f_{n,n'} \\
    \vdots \\
  g_{N_z,n'} \\
  f_{N_z,n'} \\
\end{array}%
\right) , \mathbf u =\left(%
\begin{array}{c}
  2 \left< S^z_1\right>\delta_{1,n'}\\
  0 \\
  \vdots \\
  2 \left< S^z_{N_z}\right>\delta_{N_z,n'}\\
  0 \\
\end{array}%
\right) , \label{eq:HGMatrixgu}
\end{equation}
where,  taking $\hbar=1$ hereafter,
\begin{widetext}
\begin{equation}
\mathbf M\left(\omega\right) = \left(%
\begin{array}{cccccccccccc}
  \omega+A_1&0    & B^+_1    & C^+_1& D_1^+& E_1^+& 0&0&0&0&0&0\\
   0    & \omega-A_1  & -C^+_1 & -B^+_1 &-E_1^+&-D_1^+&0&0&0&0&0&0\\
   \cdots & \cdots & \cdots &\cdots&\cdots&\cdots&\cdots&\cdots&\cdots&\cdots&\cdots&\cdots\\
 \cdots&D_n^-&E_n^-&B^-_{n}&C^-_{n}&\omega+A_{n}&0&B^+_{n}&C^+_{n}&D_n^+&E_n^+&\cdots\\
 \cdots&-E_n^-&-D_n^-&-C^-_{n}&-B^-_{n}&0&\omega-A_{n}&-C^+_{n}&-B^+_{n}&-E_n^+&-D_n^+&\cdots\\
         \cdots  & \cdots & \cdots & \cdots &\cdots&\cdots&\cdots&\cdots&\cdots&\cdots&\cdots&\cdots \\
  0& 0&0&0& 0& 0& D^-_{N_z}& E^-_{N_z}  & B^-_{N_z}   & C^-_{N_z}   &\omega + A_{N_z}&0\\
  0&0&0&0&0&0&-E^-_{N_z}& -D^-_{N_z} & -C^-_{N_z}  & -B^-_{N_z}&0  & \omega-A_{N_z}\\
\end{array}%
\right) \label{eq:HGMatrixM}
\end{equation}
\end{widetext}
where
\begin{eqnarray}
A_{n} &=& - 8J_1^{//}<S^z_n>(1+d_n-\gamma)\nonumber\\
&&-2 <S^z_{n+1}>\cos\theta_{n,n+1}(d_n+J_1^\bot)\nonumber\\
&&-2 <S^z_{n-1}>\cos\theta_{n,n-1}(d_n+J_1^\bot)\nonumber\\
&&-2J_2 < S^z_{n+2}>\cos\theta_{n,n+2}\nonumber\\
&&-2J_2 < S^z_{n-2}>\cos\theta_{n,n-2}
\end{eqnarray}
where $n=1,2,...,N_z$, $d_n=I_1/J_1^\bot$, and
\begin{eqnarray}
B_n^\pm &=& 2J_1^\bot \left< S^z_{n}\right>(\cos\theta_{n,n\pm 1}+1) \nonumber\\
C_n^\pm &=& 2J_1^\bot \left< S^z_{n}\right>(\cos\theta_{n,n\pm 1}-1)\nonumber\\
E_n^\pm &=& J_2 \left< S^z_{n}\right>(\cos\theta_{n,n\pm 2}-1)\nonumber\\
D_n^\pm &=& J_2 \left< S^z_{n}\right>(\cos\theta_{n,n\pm 2}+1) \nonumber
\end{eqnarray}
Note that to use the above formulas, we have to apply the following rules: (i) if $n=1$ then there are no  $n-1$ and $n-2$ terms in the matrix coefficients, (ii) if $n=2$ then there are no $n-2$ terms, (iii) if $n=N_z$ then there are no  $n+1$ and $n+2$ terms, (iv) if $n=N_z-1$ then there are no $n+2$ terms.  Besides, we have distinguished the in-plane NN interaction $J_1^{//}$ from the inter-plane NN one $J_1^\bot$.

\section{Results and Discussion}

Using the spectral theorem which relates the correlation
function \(\langle S^-_i S^+_j\rangle \) to the Green's function \cite{Diep2015}, we have
\begin{eqnarray}
\left< S^-_i S^+_j\right> &=& \lim_{\varepsilon\rightarrow 0}
\frac{1}{\Delta}\int\int d\mathbf k_{xy}
\int^{+\infty}_{-\infty}\frac{i}{2\pi}\big( g_{n, n'}\left(\omega
+ i\varepsilon\right)\nonumber\\
&-& g_{n, n'}\left(\omega - i\varepsilon\right)\big)
\frac{d\omega}{e^{\beta\omega} - 1}e^{i\mathbf
k_{xy}\cdot\left(\mathbf R_i -\mathbf R_j\right)},
\end{eqnarray}
where $\epsilon$ is an  infinitesimal positive constant and
$\beta=(k_BT)^{-1}$, $k_B$ being the Boltzmann constant.
Using the Green's function presented above, we can calculate
self-consistently various physical quantities as functions of
temperature $T$.  The magnetization $\langle S_{n}^z\rangle$ of the $n$-th layer is given by
\begin{eqnarray}
\langle S_{n}^z\rangle&=&\frac{1}{2}-\left< S^-_{n} S^+_{n}\right>\nonumber\\
&=&\frac{1}{2}-
   \lim_{\epsilon\to 0}\frac{1}{\Delta}
   \int
   \int d{\bf k_{xy}}
   \int\limits_{-\infty}^{+\infty}\frac{i}{2\pi}
   [ g_{n,n}(\omega+i\epsilon)\nonumber\\
   &&-g_{n,n}(\omega-i\epsilon)]
\frac{d\omega}{\mbox{e}^{\beta \omega}-1}\label{lm1}
\end{eqnarray}
After some steps,
we obtain \cite{Diep2015}
\begin{equation}\label{lm2}
\langle S_{n}^z\rangle=\frac{1}{2}-
   \frac{1}{\Delta}
   \int
   \int dk_xdk_y
   \sum_{i=1}^{2N_z}\frac{D_{2n-1}(\omega_i)}
   {\mbox{e}^{\beta \omega_i}-1}
\end{equation}
where $n=1,...,N_z$, and $D_{2n-1}(\omega_i)$ is the determinant obtained by replacing the $(2n-1)$-th column of
$\mathbf M$ by $\mathbf u$ at $\omega_i$.
As $<S_{n}^z>$ depends on the magnetizations of the neighboring layers via $\omega_i (i=1,...,2N_z)$,
we should solve by iteration the equations
(\ref{lm2}) written for all layers, namely for  $n=1,...,N_z$, to obtain the magnetizations of layers 1, 2, 3, ..., $N_z$
at a given temperature $T$. Note that by symmetry, $<S_1^z>=<S_{N_z}^z>$, $<S_{2}^z>=<S_{N_z-1}^z>$, $<S_3^z>=<S_{N_z-2}^z>$, and so on.
Thus, only $N_z/2$ self-consistent layer magnetizations are to be calculated.

The value of the spin in the layer $n$ at $T=0$ is calculated by

\begin{equation}\label{surf38}
\langle S_{n}^z\rangle(T=0)=\frac{1}{2}+
   \frac{1}{\Delta} \int \int dk_xdk_y
   \sum_{i=1}^{N_z}D_{2n-1}(\omega_i)
\end{equation}
where the sum is performed over $N_z$ negative values of  $\omega_i$ (for positive values the Bose-Einstein factor is equal to 0 at $T=0$).

The transition temperature $T_c$ can be calculated in a self-consistent manner by iteration, letting all  $<S_{n}^z>$  tend to zero, namely $\omega_i\rightarrow 0$. Expanding $\mbox{e}^{\beta \omega_i}-1\rightarrow  \beta_c \omega_i$ on the right-hand side of Eq. (\ref{lm2}) where $\beta_c=(k_BT_c)^{-1}$, we have by putting $\langle S_{n}^z\rangle=0$ on the left-hand side,
\begin{equation}\label{tcc}
\beta_c=2\frac{1}{\Delta}\int \int dk_xdk_y
   \sum_{i=1}^{2N_z}\frac{D_{2n-1}(\omega_i)}{\omega_i}
\end{equation}
There are $N_z$ such equations using Eq. (\ref{lm2}) with $n=1,...,N_z$.  Since the layer magnetizations tend to zero at the transition temperature from different values, it is obvious that we have to look for a convergence of the solutions of the equations Eq. (\ref{tcc}) to a single value of $T_c$.

\subsection{Results}

Let us take $J_1^\bot=J_1^{//}=J=1$ everywhere except on the surface where $J_1^{//}=J_s$. We use $d=I_{i,j}/J$ for any NN pair, for simplicity.

 Numerically, we use a Brillouin zone of $100^2$ wave-vector values, and we use the obtained values $\langle S_{n}^z\rangle$ at  a given $T$ as input for a neighboring $T$. At low $T$ and up to $\sim \frac{3}{5} T_c$, only a few iterations suffice to get a convergence precision $\leq 1\%$. Near $T_c$, the convergence is much harder.  We show below our results.

\begin{figure}[htb]
\centering
\includegraphics[width=7cm,angle=0]{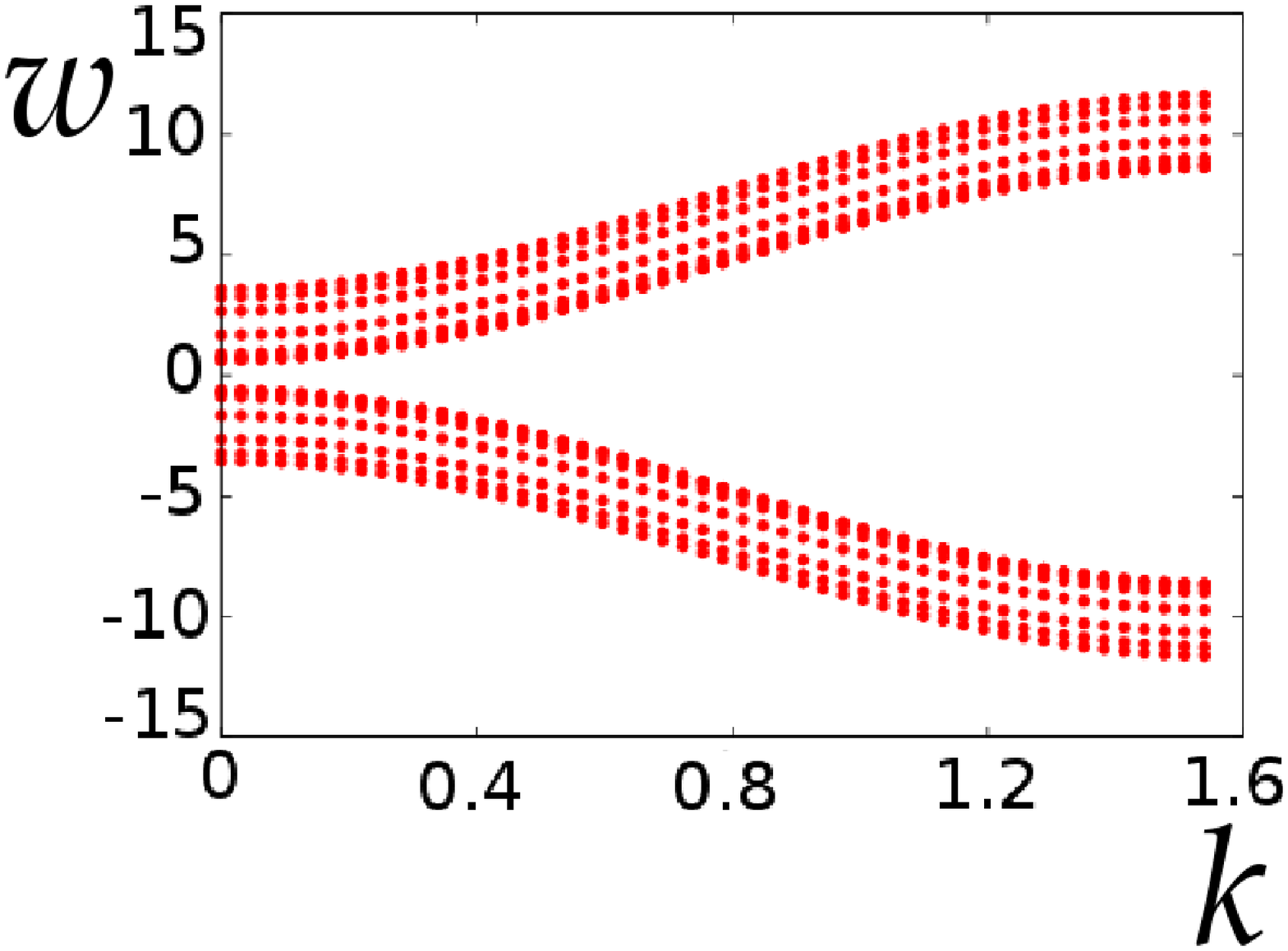}
\includegraphics[width=7cm,angle=0]{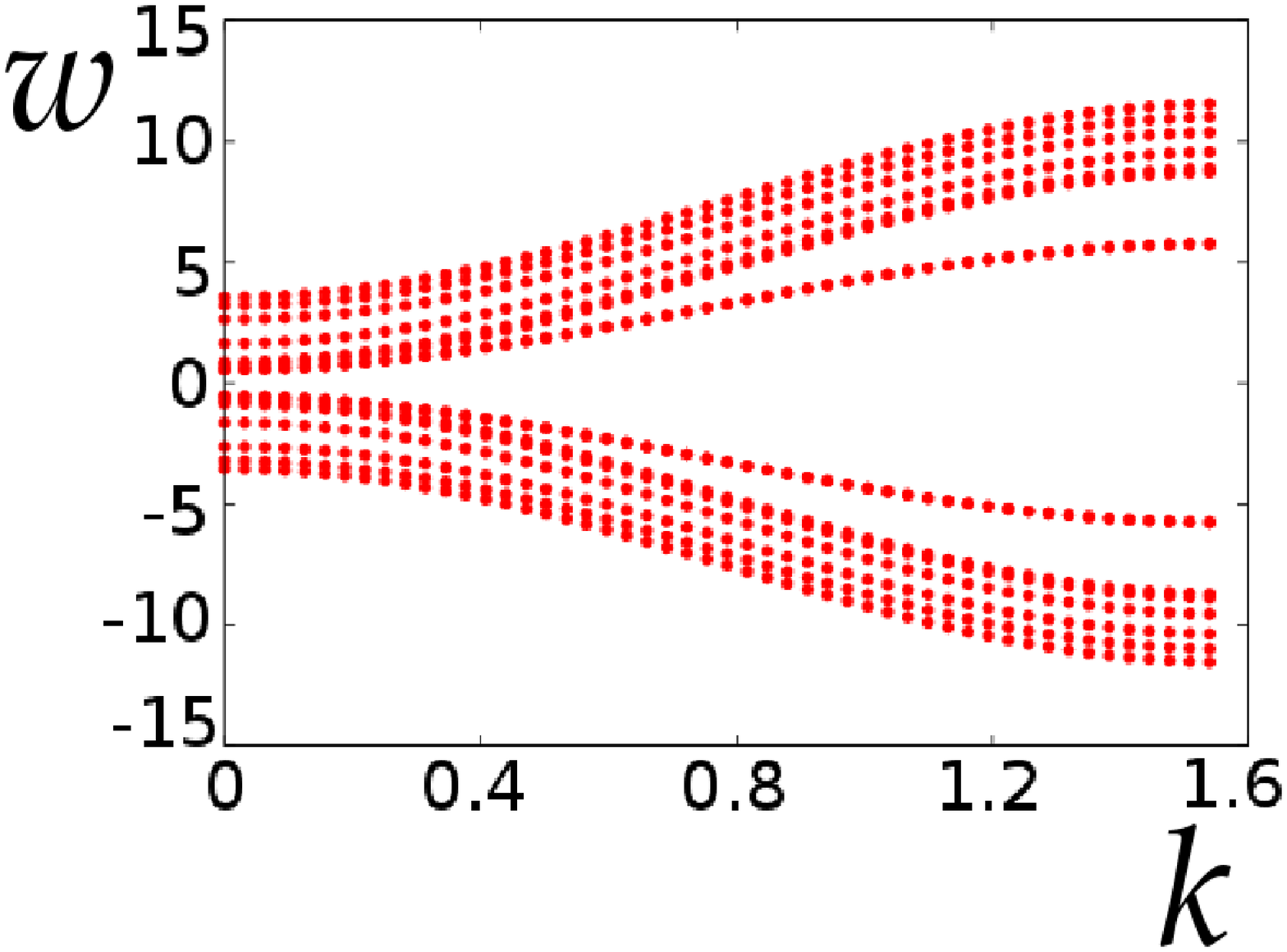}
\includegraphics[width=7cm,angle=0]{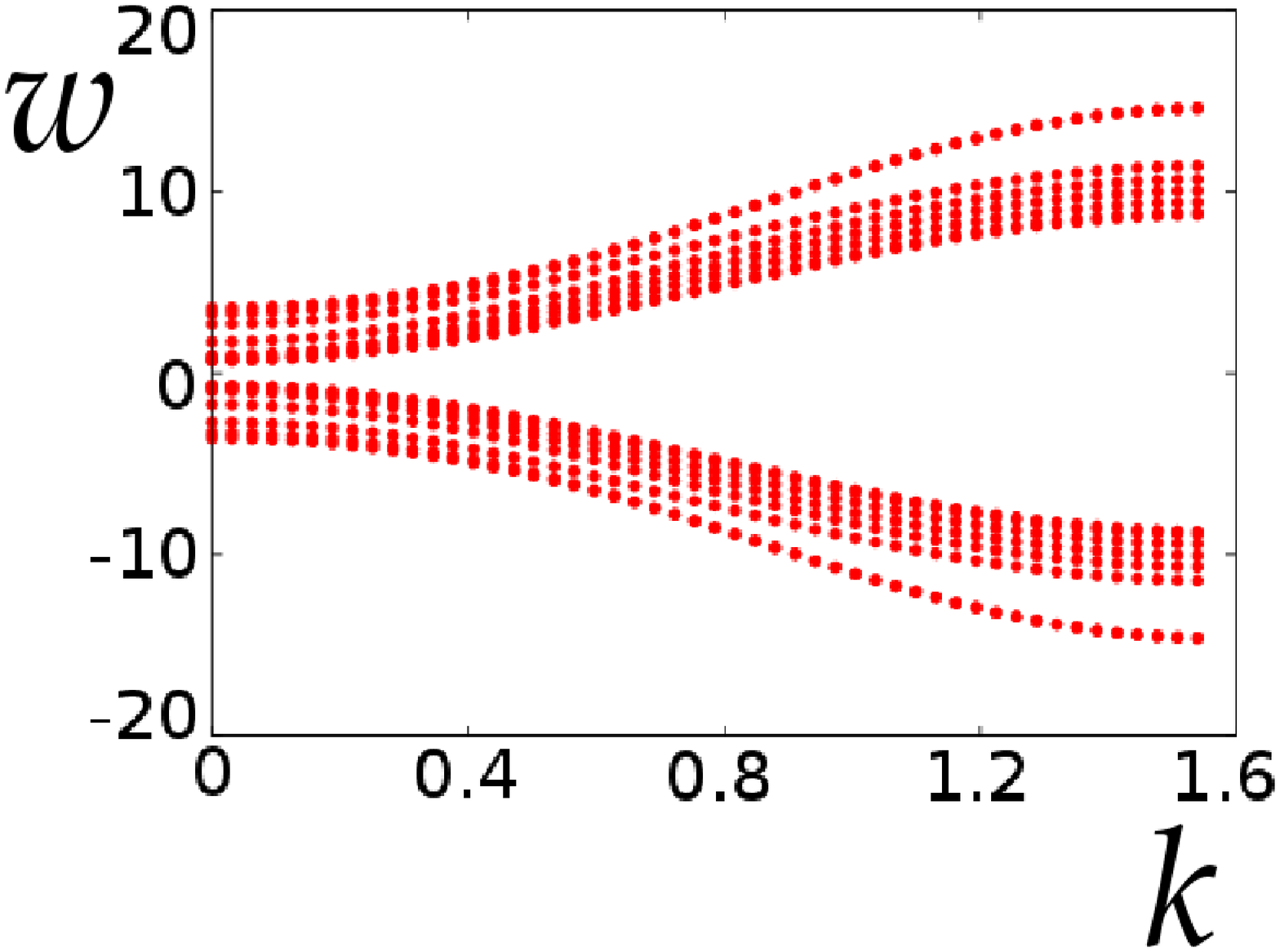}
\caption{(Color online) Spin-wave spectrum versus $k\equiv k_x=k_y$ in the case where $N_z=8$
and $d=0.1$ for $J_s=1$ (top), $J_s=0.6$ (middle) and $J_s=1.6$ (bottom).}\label{sw}
\end{figure}

We have calculated the spin-wave spectrum $\omega$ versus $k_x=k_y$
for various values of $J_2$  in the case of a
eight-layer film with an anisotropy $d=0.1$. There are 8 positive and 8 negative modes corresponding two opposite spin precessions. We can mention here the existence of acoustic
surface modes which lie in the low energy region for $J_s=0.6$ as seen in Fig.\ref{sw} (middle) and
optical surface branches which lie outside the bulk-mode energy region for $J_s=1.6$ seen in Fig.\ref{sw} (bottom),
whereas no such modes exist in the case when $J_s=1$ [Fig.\ref{sw} (top)].

\begin{figure}[htb]
\centering
\includegraphics[width=7cm,angle=0]{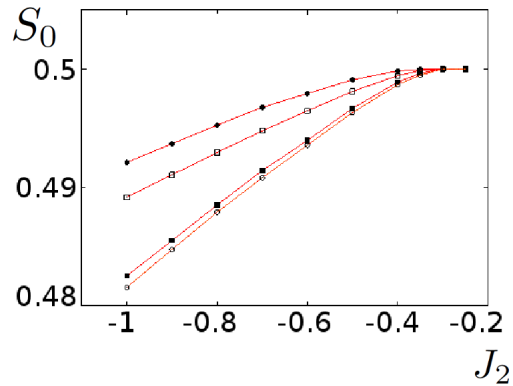}
\caption{(Color online) Spin lengths at $T=0$ for several values of $J_2$ with $d=0.1$, $N_z=8$: black circles, void squares, black squares and void circles are data for spins in first, second, third and fourth layers, respectively. }\label{contraction}
\end{figure}

It is known that in antiferromagnets, quantum fluctuations give rise to a contraction of the spin length at zero temperature \cite{DiepTM}.  We will see here that a spin under a stronger antiferromagnetic interaction has a stronger zero-point spin contraction. The spins near the surface serve for such a test. In the case of the film considered above, spins in the first and in the second layers have only one antiferromagnetic NNN while interior spins have two NNN, so the contraction at a given $J_2/J_1$ is expected to be stronger for interior spins. This is verified with the results shown in Fig. \ref{contraction}.   When $|J_2|/J_1$ increases, namely the antiferromagnetic interaction becomes stronger, we observe  stronger contractions. Note that the contraction tends to zero when the spin configuration becomes ferromagnetic, namely $J_2$ tends to -0.25.

\begin{figure}[htb]
\centering
\includegraphics[width=7cm,angle=0]{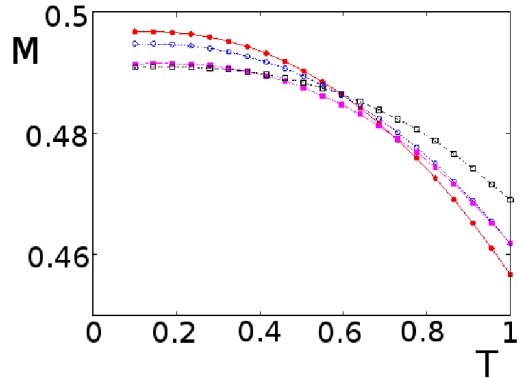}
\caption{(Color online) Layer magnetization as function of $T$ for $J_2=-0.7$ with $d=0.1$, $N_z=8$: red circles, blue void circles, magenta squares and black void squares are magnetizations of the first, second, third and fourth layers, respectively.}\label{lmc}
\end{figure}

\begin{figure}[htb]
\centering
\includegraphics[width=7cm,angle=0]{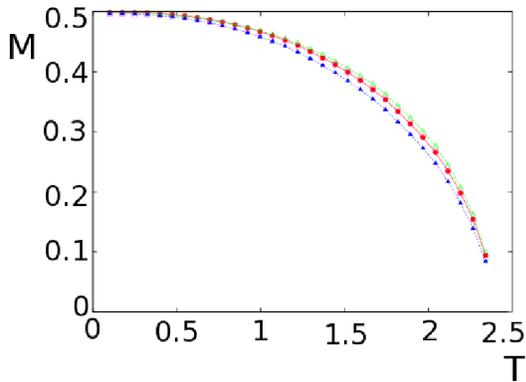}
  \caption{(Color online) Layer magnetizations as function of $T$ for $J_2=-0.5$ with $d=0.1$,$d_s=0.2$, $N_z=8$: red circles, green void triangles, blue triangles and magenta circles are magnetizations of the first, second, third and fourth layers, respectively.}\label{lma}
\end{figure}

\begin{figure}[htb]
\centering
\includegraphics[width=7cm,angle=0]{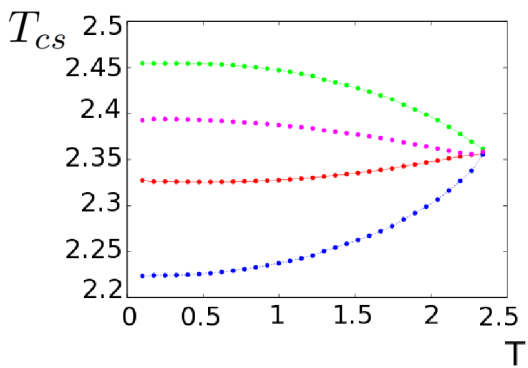}
 \caption{(Color online) Example of how to determine the transition temperature for $J_2=-0.5$ with $d=0.1$, $d_s=0.2$, $N_z=8$ (see text): red, green, blue and magenta circles are $T_{cs}$ determined from formula (\ref{tcc}) for $n=1,2,3,4$, respectively.  Their intersection gives the transition temperature $T_c\simeq 2.33$.}\label{tc}
\end{figure}


We show the layer magnetizations in Fig. \ref{lmc} in the case where  $J_2/J=-0.7$ and $N_z=8$.  Some remarks are in order:

(i) the shown result is obtained with a convergence of $1\%$. For temperatures closer to the transition temperature $T_c$, we have to lower the precision to a few percents which reduces the clarity because of their close values (not shown).

(ii) the surface magnetization, which has a large value at $T=0$ as seen in Fig. \ref{contraction}, crosses the interior layer magnetizations at $T\simeq 0.6$ to become smaller than interior magnetizations at higher temperatures.  This cross-over phenomenon is due to the competition between quantum fluctuations, which dominate low-$T$ behavior, and the low-lying surface spin-wave modes which strongly diminish the surface magnetization at higher $T$.  Note that the second-layer magnetization makes also a crossover at $T\simeq 0.6$.  Similar cross-overs have been observed in quantum antiferromagnetic films \cite{DiepTF91} and quantum superlattices \cite{DiepSL89}.

Note that though the layer magnetizations are different at low temperatures, they will tend to zero at a unique transition temperature as seen below.  The reason is that as long as an interior layer magnetization is not zero, it will act on the surface spins as an external field, preventing them to become zero.

Let us show  in Fig. \ref{lma} another example of layer magnetizations (without zoom at low $T$) up to temperatures close to the transition, for $J_2=-0.5$. The convergence is rather good but it is difficult to get to $T_c$. We explain how to determine $T_c$ by another way which is easier. As said earlier, each equation (\ref{tcc}) for a given $n$ gives a pseudo transition temperature $T_{cs}$ as long as $T$ is not close to the temperature where all layer magnetizations are very small. To determine this temperature, we plot $T_{cs}$ obtained at several temperatures. The convergence of these temperatures to a single one occurs when $T=T_c$. This is shown in Fig. \ref{tc}.

\subsection{Discussion}
Let us compare the results found in this paper for a thin film of sc lattice and those for a thin film of bcc lattice studied in Ref. \onlinecite{Diep2015}:

(i) both represent a strong non uniform spin reconstruction as a function of $J_2$.  Note that the critical value $J_2^c$ is $-0.25J_1$ in the sc case while it is $-J_1$ in the bcc case. So, the angle variation at the surface is not the same for a given value of $J_2$ in the two cases.

(ii) both show a cross-over of layer magnetizations at low temperatures, however the order of the layer magnetizations before as well as after the cross-over is not the same in the two cases.

(iii) the zero-point spin contraction is different in two cases: the sc case shows the first-layer spin contracts less than the second, the second less than the third, the third less than the fourth (see Fig. \ref{contraction}), while in the bcc case the fourth layer contracts less than the others (see Fig. 4 of Ref. \onlinecite{Diep2015}). This is in agreement with the spin contractions discussed in point (ii) above and can be understood by looking at the antiferromagnetic contribution to the local field at a spin of each layer: the smaller this contribution the smaller the contraction.  Besides, the bcc spins contract more strongly than the sc ones.

(iv) the spin-wave spectrum is different in the two cases: in the case where surface interactions are the same as the bulk interactions, the sc spectrum does not have surface-localized spin wave while the bcc spectrum has an acoustic surface branch very similar to the antiferromagnetic cases shown in Ref. \onlinecite{Diep1979}). This is because the surface spins lack four NN while the sc spins lack only one NN.  When surface interactions are smaller (larger) than the bulk ones the sc shows acoustic (optical) surface modes (see Fig. \ref{sw}). The bcc case shows similar effects but at different values of $J_2^s$.

The above qualitative and quantitative similarities and differences are very important when one deals either theoretically or experimentally with the films of different lattice symmetries.

 \section{Conclusion}\label{conclu}

 Surface effects in a helimagnet of simple cubic lattice with quantum Heisenberg spins have been investigated in this paper starting from the classical ground-state spin configuration which is exactly determined. The strong surface spin rearrangement is observed but it is insensitive to the film thickness in agreement with experiments performed on MnSi films \cite{Karhu2011} and holmium \cite{Leiner}.   We have calculated self-consistently physical quantities such as the spin-wave excitation, the spin length at $T=0$ and the layer magnetizations as functions of temperature.  We have shown that when varying the surface exchange interaction, we observe surface-localized acoustic and optical modes which lie outside the propagating-magnon energy band. These modes cause a strong deviation of the surface magnetization with respect to the interior ones.  Another interesting phenomenon is the cross-over of layer magnetizations at low temperatures due to the competition between quantum fluctuations and thermal effects.  A comparison of the results found here with those for the bcc case \cite{Diep2015} has been given. 

\acknowledgments
SEH acknowledges  a financial support from Agence Universitaire de la Francophonie (AUF).

{}

\end{document}